\documentclass[aps,prl,twocolumn,showpacs]{revtex4}

\usepackage{graphicx}

\begin{document}

\title{Resilience to damage of graphs with degree
correlations}

\author{Alexei V\'azquez$^1$ and Yamir Moreno$^2$}

\affiliation{$^1$INFN and International School for Advanced Studies, Via
Beirut 4, 34014 Trieste, Italy}

\affiliation{$^2$The Abdus Salam International Centre for Theoretical
Physics, Condensed Matter Group, P.O.Box 586, Trieste I-34014, Italy.}

\date{\today}

\begin{abstract}

The existence or not of a percolation threshold on power law
correlated graphs is a fundamental question for which a general
criterion is lacking. In this work we investigate the problems of site
and bond percolation on graphs with degree correlations and their
connection with spreading phenomena.  We obtain some general
expressions that allow the computation of the transition thresholds or
their bounds. Using these results we study the effects of assortative
and disassortative correlations on the resilience to damage of
networks.
 
\end{abstract}


\maketitle

The graphs representing many real networks are characterized by power
law degree distributions \cite{ab01a}. The origin of these power laws
can be traced back to the growing nature of real networks and to some
effective preferential attachment mechanism. This later mechanism
implies that when new vertices are added to the graph they are more
likely linked to existing vertices with large degrees
\cite{ab01a,dm02b}. Recently, there has been a great interest in the
study of processes running on top of these graphs due to their social,
technological and scientific relevance. Percolation processes
\cite{perc,ceah00}, spreading phenomena \cite{spr,yamir,n02b}, the Ising
model \cite{ising} and searching techniques \cite{search} are some
examples for which analytical solutions have been found in random
graphs with the only constraint given by the degree distribution. One
of the fundamental results is that the threshold characterizing the
percolation transition or an epidemic outbreak, depends on the ratio
$\left<d^2\right>/\left<d\right>$ of the first two moments of the
degree distribution \cite{perc,ceah00,spr,ising}.  Hence, if
$\left<d^2\right>$ diverges when increasing the graph size, there is
no transition in the thermodynamic limit.

The topology of real networks is also characterized by degree
correlations \cite{pvv01,n02a} and, therefore, the extension of
previous results for uncorrelated graphs is of utmost importance.
Moreover, it has been shown that growing network models with
\cite{kr01} and without \cite{chk01} preferential attachment lead to
non-trivial degree correlations. The study of models on graphs with
degree correlations is quite recent \cite{n02a,bp02,bl02,vw02}. Some
expressions for the size of the giant component and related quantities
have been obtained in Ref. \cite{n02a} whereas an equation for the
epidemic threshold has been provided in \cite{bp02}. General
statistical mechanics approaches for models on correlated graphs has
also been developed in Refs. \cite{bl02,vw02}.  However, in contrast
to the case of uncorrelated graphs, a general criterion for the
existence or not of a transition threshold has not been proposed
yet. A first step in this direction has been taken in
Ref. \cite{bpv02} for a disease spreading model.

In this paper we study the resilience to damage (vertex or edge
removal) of random graphs with arbitrary degree distributions and
correlations by addressing the problem of dilute (site or bond)
percolation on these graphs. We report a general equation for the
threshold and bound it. Besides, we analyze the
effect of correlations considering some examples of uncorrelated,
assortative and disassortative correlated graphs or their mixture.  We
conclude that assortative correlations can make graphs quite robust,
even with a finite $\left<d^2\right>$. On the contrary, disassortative
correlations can make graphs fragile, even with a divergent
$\left<d^2\right>$.

Let us start by considering the set of undirected graphs with $N$
vertices and arbitrary degree distribution $p_d$. Following one end of
a randomly chosen edge, we will find a vertex of degree $d$ with
probability $q_d=dp_d/\left<d\right>$.  We further assume correlations
between adjacent vertices: The conditional probability $p(d'|d)$ that
a vertex of degree $d'$ is reached following any edge coming from a
vertex of degree $d$ explicitly depends on both $d$ and
$d'$. Consistency with the degree distribution requires
$\sum_{d'}p(d'|d) = 1$. Besides, the joint probability $p(d'|d)q_d$
that the two vertices at either end of a randomly chosen edge have
degrees $d$ and $d'$ must be symmetric. For uncorrelated networks
$p(d'|d)=q_{d'}$ that is independent of $d$.

The problem of percolation on graphs with degree correlations has
been recently studied \cite{n02a} using the generating function
formalism. Alternatively, one can use a more general
statistical mechanics approach \cite{vw02}.  In this
case the size of the giant component, the fraction of nodes in the
largest cluster, is given by
\begin{equation}
S = 1 - \sum_d p_d (u_d)^d,
\label{Sperc}
\end{equation}
\begin{equation}
u_d = \sum_{d'} p(d'|d) (u_{d'})^{d'-1},
\label{piperc}
\end{equation}
where $u_d$ is the average probability that an edge connected to a 
vertex of degree $d$ leads to another vertex that does not belong to 
the giant component \cite{n02a}.

Let us generalize this result to the site
percolation problem. In this case a fraction $f$ of the nodes is
removed from the graph and the new giant component is computed.  Since
the node removal is independent of the node degree this is equivalent
to replace the original degree distribution and correlations by: $(i)$
the probability that a node selected at random has degree $d$ and it
has not been removed, and $(ii)$ the probability that if we select a
node at random and follow one of its edges we end in a node with
degree $d'$ that has not been removed, {\em i.e.}
\begin{equation}
p_d \rightarrow (1-f)p_d,\ \ \ \ 
p(d'|d) \rightarrow (1-f)p(d'|d),
\label{siteperc}
\end{equation}
Substituting Eqs. (\ref{siteperc}) in Eqs. (\ref{Sperc}) and
(\ref{piperc}) we get
\begin{equation}
S = 1 - f - (1-f)\sum_dp_d(u_d)^d,
\label{Ssite}
\end{equation}
\begin{equation}
u_d = f + (1-f)\sum_{d'}p(d'|d)(u_{d'})^{d'-1},
\label{pisite}
\end{equation}
where the term $-f$ ($f$) in Eq. (\ref{Ssite}) ( Eq. (\ref{pisite})) gives the probability of
hitting a removed node.  One solution to these equations is $u_d=1$
yielding $S=0$. This solution is valid whenever the equation for the
$u_d$ is stable under successive approximations. That is, if we start
with $u_d(n)=1-\rho_d(n)$ and compute the successive approximation
$\rho_d(n+1)$ then we should obtain that $\rho_d(n)\rightarrow0$ in
the limit $t\rightarrow\infty$. For $\rho_d(n)\ll1$ the last equation
is approximated by the linear map
\begin{equation}
\rho_d(n+1) = \sum_{d'}L_{dd'}\rho_{d'}(n),
\label{map}
\end{equation}  
with
\begin{equation}
L_{dd'} = (1-f)C_{dd'},\ \ \ \ 
C_{dd'} = (d'-1)p(d'|d).
\label{ope}
\end{equation}

The stability of the solution $u_d=1$ is then related to the largest
eigenvalue of $L_{dd'}$. If it is smaller (larger) than 1 the
solution is stable (unstable). Since $L_{dd'}$ is linear in $f$ the
stability condition can be written as
\begin{equation}
f>f_c,\ \ \ \ (1 - f_c) \Lambda_{max} = 1,
\label{threshold}
\end{equation}
where $\Lambda_{max}$ is the largest eigenvalue of $C_{dd'}$ provided
that $\Lambda_{max}>1$. If $\Lambda_{max}<1$ the graph does not
have a giant component even for $f=0$.  Moreover, since $C_{dd'}$ is
a positive matrix then $\Lambda_{max}$ has the lower and upper
bounds $\mbox{min}_d\sum_{d'}C_{dd'}$ and
$\mbox{max}_d\sum_{d'}C_{dd'}$, yielding
\begin{equation}
\mbox{min}_d \left<d\right>_{nn}(d) \leq 1+\Lambda_{max} \leq \mbox{max}_d \left<d\right>_{nn}(d).
\label{bounds}
\end{equation}
where 
\begin{equation}
\left<d\right>_{nn}(d) = \sum_{d'}p(d'|d)d',
\label{Cknn}
\end{equation}
is the average degree among the neighbors of a node with degree $d$
\cite{pvv01}. Eq. (\ref{bounds}) can be used to determine, based on a
simple topological measure, whether or not a given graph is robust
under vertex removal.

In the bond percolation problem a fraction $f$
of the edges is removed from the graph and the new giant component is
computed.  Since the edge removal is made at random this is equivalent
to keep the original degree distribution and replace the degree
correlations by the probability that if we select a node at random and
follow one of its edges we end in a node with degree $d'$ that has
not been removed, {\em i.e.}
\begin{equation}
p_d \rightarrow p_d,\ \ \ \ 
p(d'|d) \rightarrow (1-f)p(d'|d).
\label{bondperc}
\end{equation}
Substitution of Eq. (\ref{bondperc}) in Eqs. (\ref{Sperc}) and
(\ref{piperc}) yields
\begin{equation}
S = 1 - \sum_dp_d (u_d)^d,
\label{Sbond}
\end{equation}
\begin{equation}
u_d = f + (1-f)\sum_{d'}p(d'|d)(u_{d'})^{d'-1}.
\label{pibond}
\end{equation}
Note that the only difference between the site and bond percolation
problems (see Eqs. \ref{Ssite},\ref{pisite}) is the equation
for the giant component while that for $u_d$ is identical. Hence,
Eqs. (\ref{threshold}) and (\ref{bounds}) are also valid for the bond
percolation problem.

In what follows we consider some particular graphs in order to
analyze the effects of correlations. Depending on the monotony of
$\left<d\right>_{nn}(d)$ the degree correlations can be classified in:
uncorrelated if it is independent of $d$, assortative or positive if it
increases with increasing $d$ and disassortative or negative if it
decreases with decreasing $d$. A similar definition has been
introduced in Ref. \cite{n02a} using a correlation coefficient.

For random graphs with no constraint other
than the one imposed by the degree distribution we have
$p(d',d)=q_{d'}$. In this case the lower and upper bounds in
Eq. (\ref{bounds}) are equal giving for the largest eigenvalue
\begin{equation}
\Lambda^{unco}_{max} = \frac{\left<d^2\right>}{\left<d\right>}-2.
\end{equation}
Alternatively one can compute $\Lambda$ directly from the eigenvalue
problem of $C_{dd'}$. Then from Eq.  (\ref{threshold}) we obtain
$1-f_c = 1/(\left<d^2\right>/\left<d\right>-2)$ \cite{ceah00}.  Hence,
if the second moment $\left<d^2\right>$ diverges the threshold equals
1, {\em i.e.}  the network is robust under random vertex or edge
removal. Furthermore, consider the case in which the degree
correlations can be decomposed into two components
\begin{equation}
p(d'|d) = \alpha q_{d'}
+ (1-\alpha)\delta p(d'|d)
\label{pddunc}
\end{equation}
with $0<\alpha<1$ and $\delta p(d'|d)>0$ for all $(d,d'$). Varying
the parameter $\alpha$ one interpolates between the uncorrelated graphs
($\alpha=1$) and a graph with arbitrary degree correlations given by
$\delta p(d'|d)$. In this case from Eq. (\ref{bounds}) we obtain
$\Lambda_{max}\geq\alpha\Lambda^{unco}_{max}$ and, therefore, if the
network is robust for the uncorrelated case it will also be robust for
any $\alpha>0$. This immediately implies that any graph with a
divergent second moment and a finite amount of random mixing of
edges does not have a percolation  threshold.

\begin{figure}[t]
\begin{center}
%\centerline{\psfig{file=fig1.eps,width=2.5in,angle=0}}
\includegraphics[width=2.5in,angle=0]{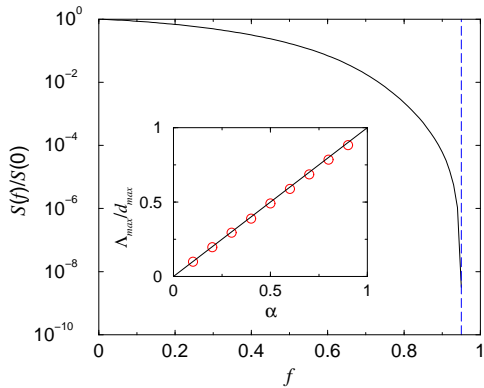}
\end{center}
\caption{Size of the giant component for a graph with $p_d=cd^{-3.5}$
($2\leq d\leq d_{max}$, $d_{max}=100$) and degree correlations
$p(d'|d) = \alpha \delta_{dd'} + (1-\alpha) q_{d'}$ with $\alpha$,
as computed from Eq. (\ref{Ssite}). The dashed line marks the
percolation threshold obtained using perturbation theory
(Eq. (\ref{Lassmix_})). The inset shows the largest eigenvalue
relative to $d_{max}$ as a function of $\alpha$. The points were
computed numerically and the line is the perturbation theory
dependency $\Lambda_{max}/d_{max}=\alpha$.}
\label{fig:1}
\end{figure}

Assortative correlations allow us to show
that the divergence of the second moment is not a necessary condition
for the absence of the threshold. Let us consider a network with
degree correlations
\begin{equation}
p(d'|d) = \alpha \delta_{dd'} + (1-\alpha)\delta p(d'|d),
\label{pddassmix}
\end{equation}
with $0<\alpha<1$ and $\delta p(d'|d)>0$ for all
$(d,d'$). $\alpha=1$ corresponds to a fully assortative graph made
up of sub-graphs with fixed degree. In this case
$C_{dd'}=d'\delta_{dd'}$ (see Eq. \ref{ope}) is already
diagonal.  The largest eigenvalue is $\Lambda_{max}=d_{max}$, where
$d_{max}$ is the largest degree. If $d_{max}$ diverges for
$N\rightarrow\infty$ then $f_c=1$. For the more general case
$0<\alpha<1$ we compute the largest eigenvalue using perturbation
theory \cite{ss90} around $\alpha=1$, obtaining
\begin{equation}
\Lambda_{max}(\alpha) = \alpha d_{max} + (1-\alpha)C_{d_{max}d_{max}}.
\label{Lassmix}
\end{equation}
This result is valid whenever $(1-\alpha)C_{d_{max}d_{max}}\ll\alpha
d_{max}$. In general $C_{d_{max}d_{max}}$ decreases with increasing
$d_{max}$, resulting
\begin{equation}
\Lambda_{max}(\alpha) \approx \alpha d_{max},
\label{Lassmix_}
\end{equation}
for $d_{max}\gg1/\alpha$. Hence, for any $\alpha>0$ and any unbounded
degree distribution we have $f_c=1$, {\em i.e.}  there is no
percolation threshold. In Fig. \ref{fig:1} we show the
validity of the perturbation theory for a particular perturbation
$\delta p(d'|d)$.  Thus, as in the fully assortative case, if
$\alpha>0$ and $d_{max}$ diverges $f_c=1$. Therefore, we can conclude
that the divergence of the second moment is not a necessary
condition. 

Let us now analyze if the divergence of the second moment is a sufficient 
condition for $f_c=1$, using an example of a
disassortative graph. Consider a vertex with degree $d>1$ and an edge
incident to it. Then with probability $g_d$ a vertex at the other end
is chosen at random among all vertices with degree $d'>1$, otherwise
it is connected to a vertex with $d'=1$ chosen at random, {\em i.e.}
\begin{eqnarray}
p(d'|d) &=& \frac{ (1-g_{d'})d'p_{d'} }{\sum_s(1-g_s)sp_s}
\Theta(d'-1)\delta_{d,1} \nonumber\\ &+&
(1-g_d)\delta_{d',1}\Theta(d-1) \nonumber\\ &+& g_d\frac{
g_{d'}d'p_{d'} }{\sum_sg_ssp_s} \Theta(d'-1)\Theta(d-1).
\label{disa}
\end{eqnarray}
where $\Theta(x)$ is the unitary step function ($\Theta(x)=0$ for
$x\leq0$ and $\Theta(x)=1$ for $x>0$).  Moreover, the fraction of
nodes with degree 1 is obtained self-consistently from the condition
$p_1=\sum_{d>1}(1-g_d)dp_d$.  The average degree of the neighbors of a node
with $d>1$ is given by
\begin{equation}
\left<d\right>_{nn} =
1+g_d\left(\frac{\sum_{d'>1}g_{d'}d'^2p_{d'}}{\sum_sg_ssp_s}-1\right),
\label{dnndisa}
\end{equation}
and, therefore, these graphs are disassortative for any monotonic
decreasing function $g_d$. To analyze the percolation properties of
this graph we computed exactly the largest eigenvalue of
$C_{dd'}=(d'-1)p(d'|d)$, resulting
\begin{equation}
\Lambda_{max} = \frac{\sum_dg_d^2(d-1)dp_d}{\sum_sg_ssp_s}.
\label{lmaxdisa}
\end{equation}
Hence, the conditions for the existence of a giant component
($\Lambda_{max}>1$) or resilience to damage ($\Lambda_{max}=\infty$)
are modulated by $g_d$ and, therefore, the disassortative correlations
given by $g_d$ have a great impact on the percolation properties. For
instance, let us consider $g_d=d^{-\alpha}$ and a power law degree
distribution $p_d=cd^{-\gamma}$ with $\gamma<3$
($\left<d^2\right>=\infty$). From Eq. (\ref{dnndisa}) it follows that
$\left<d\right>_{nn}-1\sim d^{-\alpha}$ so that when increasing
$\alpha$ the graph gets more and more disassortative. Moreover,
$\Lambda_{max}$ diverges for $\alpha<\alpha_c=(3-\gamma)/2$ and it is
finite otherwise. Thus, for small values of $\alpha$ the graph is
robust but for $\alpha>\alpha_c$ it becomes fragile. It is worth
noticing that the value of $\alpha$ above which the giant component
disappears ($\Lambda_{max}<1$) is larger than $\alpha_c$. Besides, for
large degrees, the degree distribution of the vertices in the giant
component is still a power law, but it decays slower than that of the whole graph.  
Thus, disassortative correlations competes against the
formation of the giant component and, the divergence of
$\left<d^2\right>$ is not a sufficient condition to get a robust
graph with $f_c=1$.

The connection between percolation theory and
models of epidemic spreading is well known \cite{grass}.  Two general
classes of epidemiological models can be related to percolation
problems, the Susceptible-Infected-Removed (SIR) and the
Susceptible-Infected-Susceptible (SIS) classes. The SIR model assumes
that individuals can exist in three classes and that once they get
infected they can not catch the infection again.  This model can be
mapped into a bond percolation problem taking $f$ as the probability
that the disease will be transmitted from one node to another and the
size of the giant component as the size of the outbreak. Hence, all
the conclusions drawn above for the bond percolation problem can be
translated to the language of epidemic spreading for the SIR model on
top of graphs with degree correlations, extending in this way
previous studies in Refs. \cite{yamir,n02b}  for uncorrelated graphs.

On the other hand, the SIS model allows individuals to move through
the cycle of infection so that the prevalence (number of infected
individuals) attains a stationary value. The SIS model on top of
graphs with degree correlations has been recently analyzed in 
Refs. \cite{bp02,bpv02}. They obtained the epidemic threshold (the
value of $\lambda$ above which the solution with zero prevalence is
unstable) $\lambda=1/\Lambda'_{max}$, where $\Lambda'_{max}$ is the
largest eigenvalue of the matrix $C'_{dd'}=dp(d'|d)$. This approach is
quite similar to the one presented here for site percolation with the
remark that $C'_{dd'}$ is different (see Eq. (\ref{ope})). In fact, if
$y_d$ is an eigenvector of $C'_{dd'}=dp(d'|d)$ corresponding to the
eigenvalue $\Lambda'$ then $y_d/d$ is an eigenvector of
$C''_{dd'}=d'p(d'|d)$ corresponding to the same eigenvalue. This last
matrix is that of Eq. (\ref{ope}), but replacing $d'$ by
$d'-1$. However, this subtle difference makes the SIS and dilute
percolation different. We have computed the largest eigenvalue of
$C''_{dd'}$ for the disassortative graph considered above(Eq.\
 (\ref{disa})). Taking the limit $\left<d^2\right>\gg1$ one gets
\begin{equation}
\Lambda'_{max} \approx 
\frac{\sum_d(1-g_d)d^2p_d}{\sum_s(1-g_s)sp_s},
\label{lmaxdisaSIS}
\end{equation}
where $g_d$ is again a decreasing function of $d$. In this case,
independent of the form of $g_d$, the divergence of the second moment
of the degree distribution implies the divergence of
$\Lambda'_{max}$. Moreover, the same conclusion is obtained if $g_d$
is an increasing function of $d$. The conditions for the existence of
a finite prevalence in the SIS model have been recently addressed in
\cite{bpv02}, where the divergence of the second moment has been shown
to be a sufficient condition for the absence of the phase transition
in the SIS model. Nevertheless, we have shown that this conclusion
does not hold for dilute percolation. This essential
difference is rooted in the existence of an additional dimension in
the SIS model, given by the time evolution of the density of infected
sites.

In summary, we have studied the percolation problem on top of random
networks with arbitrary degree distribution and correlations, making
its generalization to site and bond percolation. The connection with
spreading phenomena was also analyzed. We provide some general
expressions to obtain or bound the transition threshold. Using these
results we have shown that the existence of a finite amount of random
mixing of the connections between vertices is sufficient to make the
graph robust under vertex or edge removal provided
$\left<d^2\right>\rightarrow\infty$.  Assortative correlations makes
the situation even better, they can lead to a graph robust to random
damage even with a finite second moment of the degree distribution. On
the contrary, disassortative correlations compete again the formation
of the giant component and can make a graph fragile even with a
divergent second moment.

We thank A.\ Vespignani, R. Pastor-Satorras and M.\ Weigt for helpful 
comments. This work has been partially supported by the European 
commission FET Open project COSIN IST-2001-33555.

\end{document}